# Growth and replication of red rain cells at 121°C and their red fluorescence

Rajkumar Gangappa[1,2], Chandra Wickramasinghe[2*], Milton Wainwright[3], A. Santhosh Kumar[4] and Godfrey Louis[4]

[1]University of Glamorgan, Trefforest, Pontypridd Wales, CF37 1DL, UK
[2]Cardiff Centre for Astrobiology, Cardiff University, Cardiff, CF10 3DY, UK
[3]Department of Molecular Biology and Biotechnology, University of Sheffield. Sheffield, S10 2TN, UK
[4] Astrobiology Division, Department of Physics, Cochin University of Science and Technology, Kochi- 682 022, India

*Corresponding author: Email: ncwick@googlemail.com, Wickramasinghe@cardiff.ac.uk

## ABSTRACT

We have shown that the red cells found in the Red Rain (which fell on Kerala, India, in 2001) survive and grow after incubation for periods of up to two hours at 121°C. Under these conditions daughter cells appear within the original mother cells and the number of cells in the samples increases with length of exposure to 121°C. No such increase in cells occurs at room temperature, suggesting that the increase in daughter cells is brought about by exposure of the Red Rain cells to high temperatures. This is an independent confirmation of results reported earlier by two of the present authors, claiming that the cells can replicate under high pressure at temperatures upto 300° C. The flourescence behaviour of the red cells is shown to be in remarkable correspondence with the extended red emission observed in the Red Rectagle planetary nebula and other galactic and extragalactic dust clouds, suggesting, though not proving an extraterrestrial origin.

## 1. INTRODUCTION

Most forms of life on Earth are adapted to growth within the temperature range, 10-45°C. Microorganisms, known as thermophiles, however, grow optimally between 45°C and 70°C. Recently novel microbes, growing at temperatures above 80°C have ben reported, and the upper temperature limit for growth has been extended to 113°C and 121°C by the discovery of *Pyrolobus fumarii* and "strain 121", both having been isolated from hydrothermal vents (Kashefi and Lovley, 2003). Stetter (1992) termed these newly discovered microorganisms "hyperthermophiles". There also exist in the literature disputed claims that bacteria, isolated from black smokers can grow at 250°C (Baross and Deming, 1983). Louis and Kumar (2003) also reported that red cells which fell in so-called "Red Rain" over at Kerala in India (and which are studied here) could replicate at 300°C (Louis and Kumar, 2003, 2006), a claim that needs to be verified independently if it is to gain acceptance.

In Part I of the present paper we report studies to determine if Red Rain cells can replicate following autoclaving for periods up to 2 hours, i.e. at a temperature of 121°C. Samples of red rain were incubated in a bacterial growth medium at 121°C for 0.5 to 2.0 hours. The number of cells present following such incubation was then determined and the cells were observed under a light microscope as well as a scanning and transmission electron microscope to determine any changes in their morphology.

Following incubation the red rain cells suspended in DMSO were also observed under phase contrast fluorescence microscopy using a range of filter types to control excitation wavelengths. These studies are reported in Part II, in which comparison with astronomical data is also discussed.

# PART 1

## 2. MATERIALS AND METHODS

### 2.1 Cell Culure

Red rain samples (500µl) were inoculated into the 5ml of sterile Luria Bertani (LB) medium (containing, sodium chloride, 10g; peptone,10g; yeast extract, 5g; ditilled water 1liter, pH-7.0). Inoculated samples were separately autocalved for 0.5, 1 hour, 1:30 hour, 2 hours at 121$^o$C. The number of cells present following this treatment was then determined using an heamocytometer (Thomas, Weber, England, depth 0.1mm, 1/400mm$^2$). Cells were counted three times separately and an average of three counts was taken in order to calculate the number of cells in each ml. Uninoculated LB medium(5ml) was included as a control.

### Transmission Electron Microscopy

The cells from each sample (before and after autoclaving) were fixed for 1 h. in glutaraldehyde (2.5% in 0.1M sodium cacodylate buffer, pH 7.5), washed twice in 0.1M sodium cacodylate buffer for 10 minutes and then fixed in osmium tetroxide(1%) for 1 h. in a refrigerator. The cells were then dehydrated on three occasions in alcohol (from 30% to 100%) followed by application of proylene oxide on each occasion, at room temperature for 10 min. The samples were then infused overnight with araldite (araldite: 5g, dodecenyl succinic anhydride: 5g, butanediol dimethacrylate: 0.15g, 1:1 mixture + propylene oxide), and then embedded for 2 days at 60$^o$C in pure araldite (araldite:5g, dodecenyl succinic anhydride: 5g, butanediol dimethacrylate: 0.15g). The resulting polymerized resin block was cut into 90-100nm thick sections on a Reichert ultracut microtome (Reichert-Jung, Austria) and then mounted on pioloform coated copper grids and stained with 2% uramyl acetate and lead citrate prior to examination. Ultrastructural examination was carried out using a Philips EM208 transmission electron microscope operated at 80KV accelerating voltage.

## 3. RESULTS AND DISCUSSION

### 3.1 Cell culture experiment

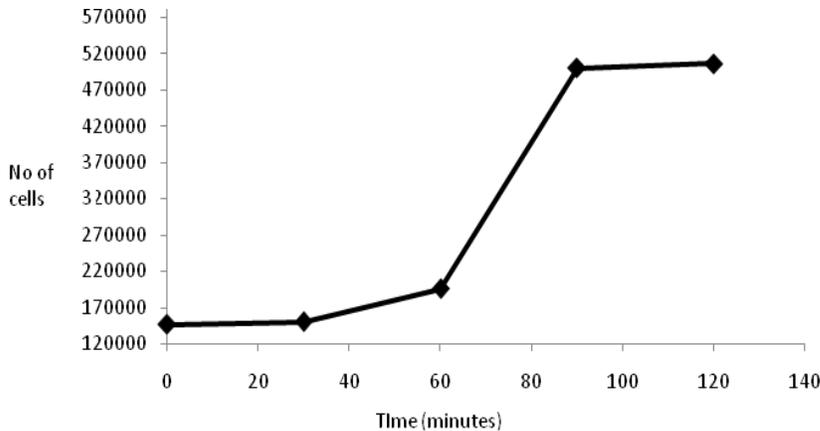

Fig.1. Number of Red Rain cells in the sample after autocaloving at 121$^o$C for various time periods.

Fig.1 shows that the number of cells in the Red Rain sample increased with increasing lengths of exposure to a temperature of 121°C, reaching a maximum after 1.5 to 2.0 hours. No similar increase occured in Red Rain when maintained at room temperature. These results show that Red Rain cells do not merely survive at 121°C, as extremodures (Wainwright,2000) but instead, they grow and replicate at this temperture, leading to an increase in the number of cells present in the sample.

## 4. MICROSCOPE STUDIES

### 4.1 Optical microscope

Red Rain cells, when kept at room temperature, are reddish-brown in colour and vary in shape from are oval to round, and of a size between 5 µm to approximatley 10µm. Pre-autoclaved cells are evenly dispersed in the rain water having a reddish centre and external dark red or brown outer envelope (Fig 2. A). Post-autoclaved cells for 1 and 2 hour, are in clumps and size varies from 1µm to approximately 10µm (Fig 2. B and C).

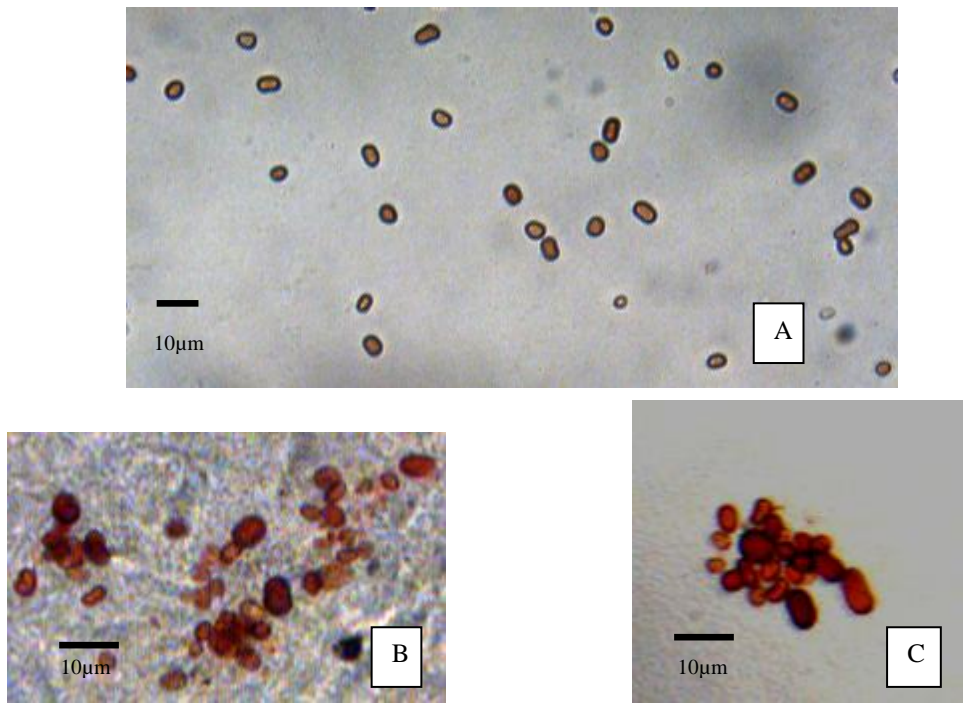

Fig 2. Optical microscope images of red cells: (A) red cells before autoclaving (400x): cells evenly dispersed in the rain water. (B) red cells after 1 hour incubation at 121°C (1000x).(C) after 2 hour incubation at 121°C (1000x).

### 4.2 Transmission Electron Microscope (TEM)

The ultrastructure of red cells before autocaving and examined by TEM shows that these cells consist of central core bounded by a double layered membrane and multilayered external envelope (Fig 3. a and b).  The external envelope is very thick and is composed of three layers. The large electron dense structure  (Fig 4. a and b, indicated by an arrow) in the central core is possibly a nucleus; further research is however, needed to clarify the nature of these structures.

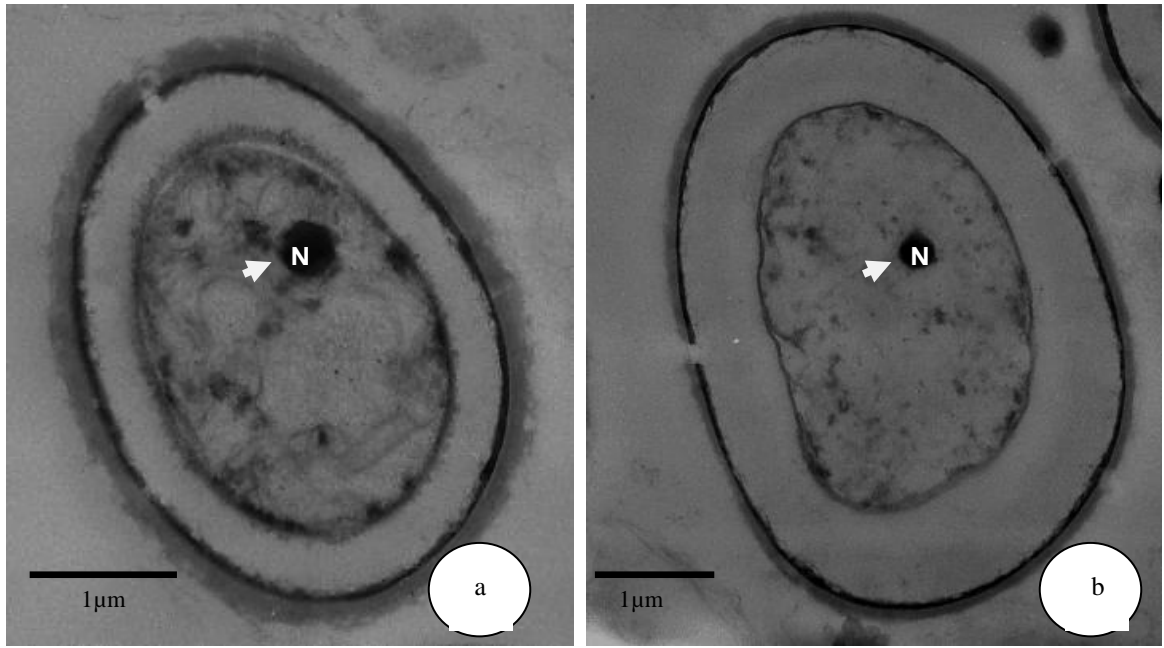

Fig.3. Transmission electron micrograph of Red Rain cells prior to autoclaving at 121°C ; (N, shown by an arrow) Possible nucleoid inside the core compartment bounded by double membrane and multilayered thick outer envelope.

Following incubation at 121°C for 1 hour and longer, a marked change occurs in the internal appearance of the Red Rain cells (Fig.4 c (i) and d (i)), as small cells appear in the original larger cells. These small cells can be regarded as "daughter cells" having the same morphology as their "mother cells". The size of the daughter cells, after 1h exposure to 121°C, ranges from 30 nm to 120 nm in size (Fig 4 c (i), (ii) and b (i), (ii)). The cell wall of these daughter cells is seen to thicken following incubation for 2hours (Fig.5 (i) and (ii)).

In conclusion, the results of the present study clearly establishes that red cells discovered in the Kerala rain, replicate at 121°C and that there is a significant increase in the number of cells after incubation at 121°C. Furthermore, optical microscopy and electron microscopy of post-incubated red cells confirms that these cells are hyperthermophiles. The formation of daughter cells having the same morphology as the mother cells clearly shows that Red Rain Cells are not single endospores, such as those seen in bacteria, such as species of *Bacillus* and *Clostridium*.

The optimum growth conditions and upper temperature limit of these cells is yet to be determined. Although autoclaving at 121°C for 20 mins kills most microorganims, some spores of *Bacillus and Clsotridium* species can resist this treatment and germinate to form vegetative cells when incubated at lower temperatures (Hyum *et* al,1983,Vessoni,*et al*.1996). Here, however, we have shown that, unlike heat resistant bacterial spores, Red Rain cells grow and produce daughter calls when incubated at 121°C for 2 hours. The results of these experiments show the remarkable ability of Red Rain cells to grow and replicate at 121°C and thereby supports the hyperthermostability of red cells, as reported by Louis and Kumar (2003); no attempt however, was made to confirm their claims that Red Rain cells grow at 300°C.

The origin of Red Rain, and the cells that it contains, has yet to be discovered, although the results of this study suggest that, since such cells are adapted to growth and reproduction at high temperatures, they likely originate in an extreme environment which is at times exposed to high temperatures; whether such environments occur on Earth, or elsewhere, has yet to be determined.

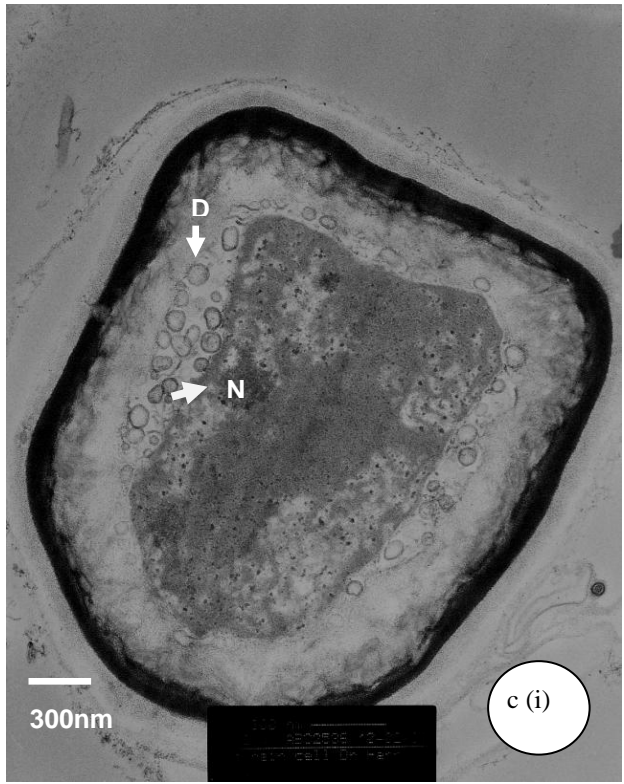
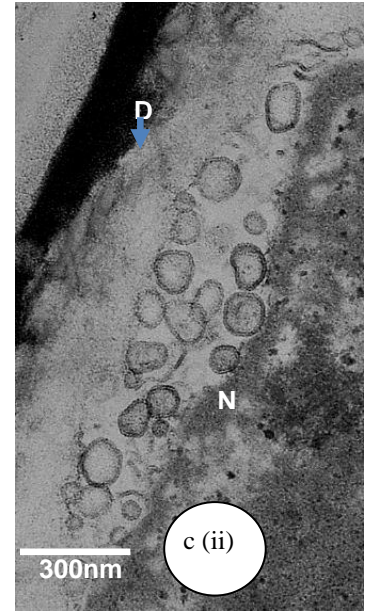
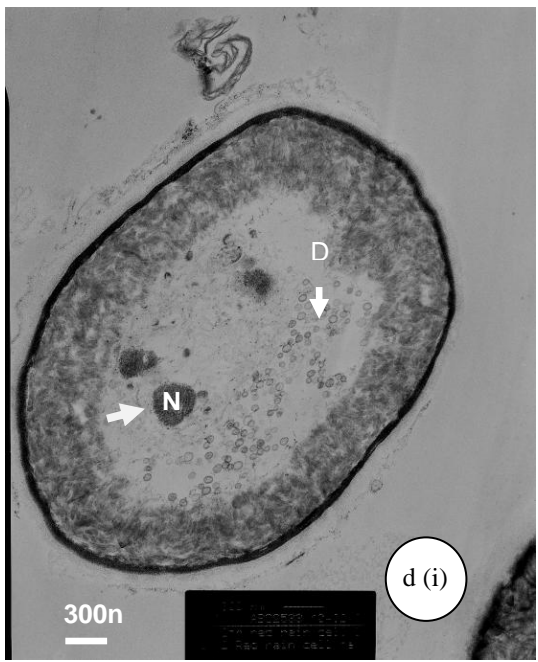
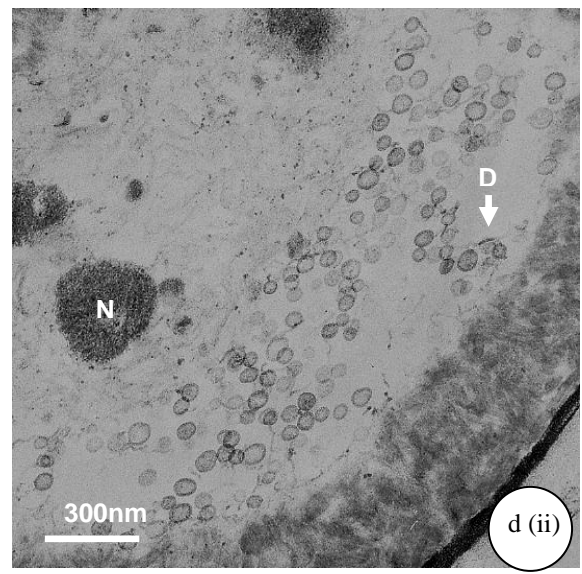

Fig.4 Transmission electron micrograph of Red Rain cells after autoclaving at 121°C for 1h showing the appearance of immature daughter cells (D, shown by blue arrow); (N, shown by an arrow) possible nucleoid structure bounded by double membrane and multilayered thick outer envelope.

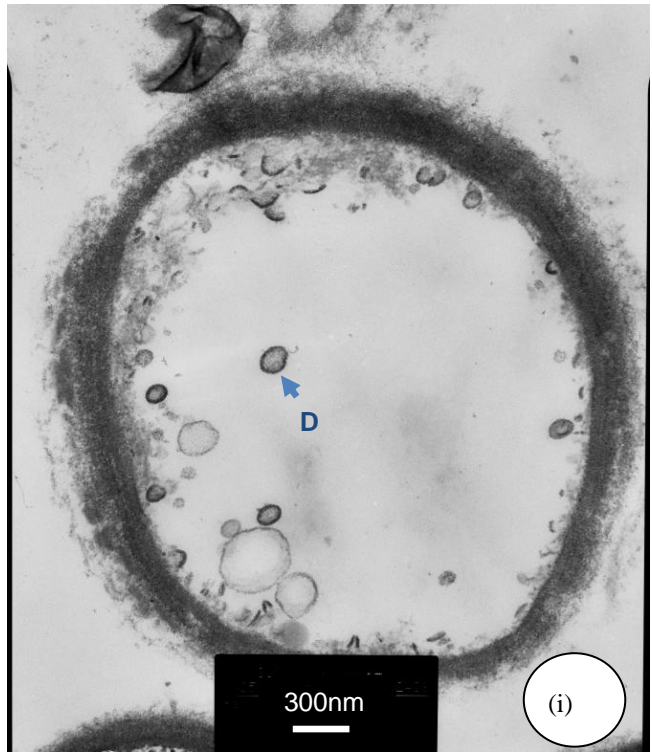

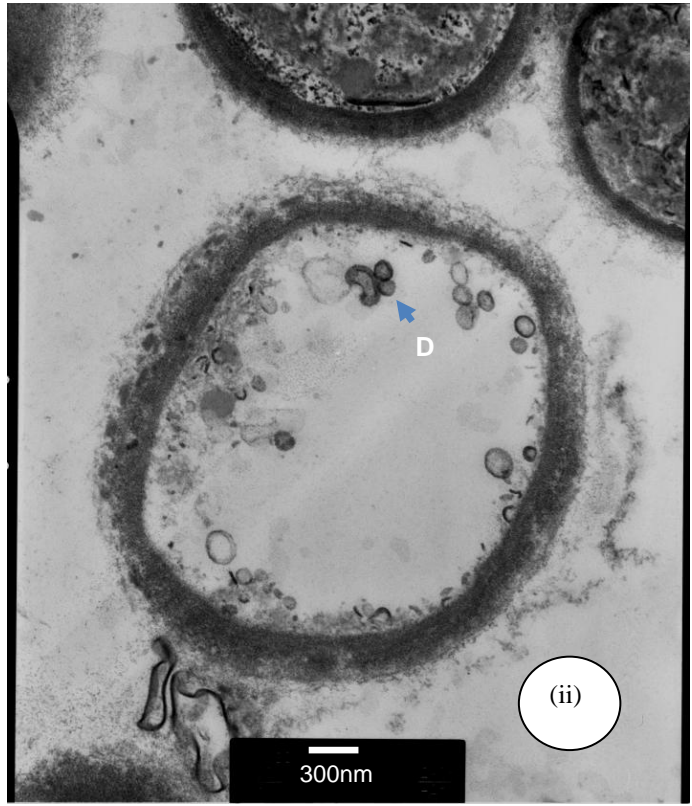

Fig.5. Transmission electron micrograph of Red Rain cells after incubating for 2h, showing the appearance of mature daughter cells (D, shown by blue arrow).

## PART II

## 5. FLOURESCENCE STUDIES

Fluorescence examination reported in this section may well have a bearing on a possible space/comet origin of the red rain cells. For fluorescence spectroscopy, 1 ml of red rain sample was centrifuged at 13000 rpm for 10 minutes and the sediment washed twice in ddH$_2$O. To the washed sediment of red rain cell 600µl of DMSO was added and incubated for 10 minutes at room temperature. After incubation red rain cell suspended in DMSO was centrifuged at 13000 rpm for 10 minutes. Red supernatant was pipetted out and its fluorescence emission measured by using AMINCO-Bowman Series 2 (AB2) Spectrofluorometer at excitation wavelength ranging from 250 nm to 600 nm.

### 5.1 Fluorescence microscopy and spectroscopy

Fluorescence examination reported in this section may well have a bearing on a possible space/comet origin of the red rain cells. For fluorescence spectroscopy, 1 ml of red rain sample was centrifuged at 13000 rpm for 10 minutes and the sediment washed twice in ddH$_2$O. To the washed sediment of red rain cell 600µl of DMSO was added and incubated for 10 minutes at room temperature. After incubation red rain cell suspended in DMSO was centrifuged at 13000 rpm for 10 minutes. Red supernatant was pipetted out and its fluorescence emission measured by using AMINCO-Bowman Series 2 (AB2) Spectrofluorometer at excitation wavelength ranging from 250 nm to 600 nm.

Images (fig.6. a–e) show the fluorescence emission pattern of red rain cells in DMSO when observed with different excitation filters. A phase contrast fluorescence microscope (Zeiss: Axiovert 10) was used with filters as defined below:

GFP (exciter 470/40 – emitter 525/50, Beam splitter T4951p),

ET/Ds Red (exciter 545/25-emitter 605/70, beam splitter T5651pxr),

YFP (exciter 450/490- emitter 520, beam splitter FT510)

CFP (exciter 395/440- emitter 470, beam splitter FT460) for fluorescence emission.

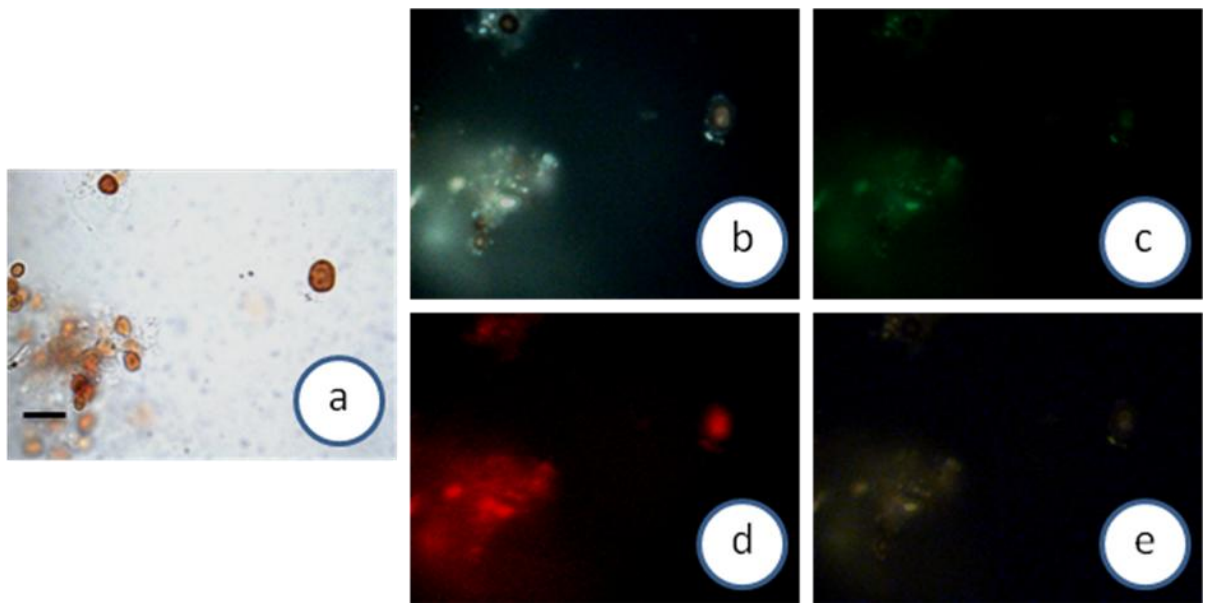

Fig. 6. Phase contrast fluorescence microscopy emission images of DMSO extract of red rain cells: (a) red rain cells suspended in DMSO, scale bar – 10 µm, 1000x magnification. (b) Fluorescence emission of red rain cells in DMSO, filter- CFP. (c) Fluorescence emission of red rain cells in DMSO, filter- GFP. (d) Fluorescence emission of red rain cells in DMSO, filter – ETS/Ds Red.

## 5.2 Fluorescence Spectroscopy

Fig. 7. (i)-(xxiii) shows the fluorescence emission spectra for the excitation wavelengths ranging from 370-420nm. The nearly invariant fluorescence behaviour for a wide range of excitation wavelengths in the blue region of the spectrum is unusual when compared with the behaviour of other biological pigments.

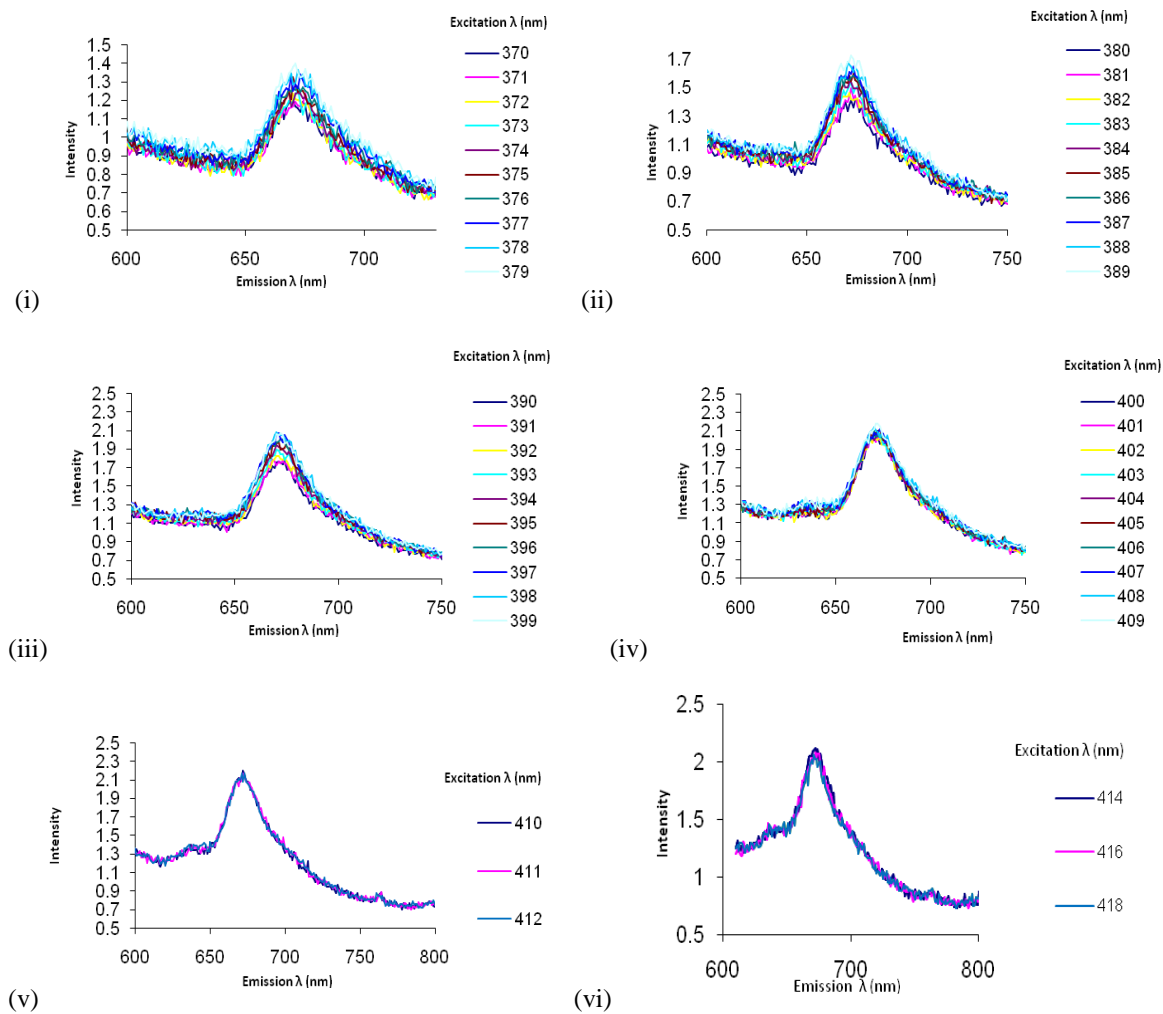

Fig. 7. Fluorescence emission spectrum of DMSO extract of red rain cells: (i – vi) red fluorescence emission showing a broad peak centred at 674 nm when excited from 370 nm to 408 nm.

More generally the emission spectra are related to the excitation spectrum, as seen for instance, in a fluorescent protein from a deep sea luminous fish studied some years ago by Campbell and Herring (1987). The excitation/emission correspondence for comparison is shown in Fig. 8.

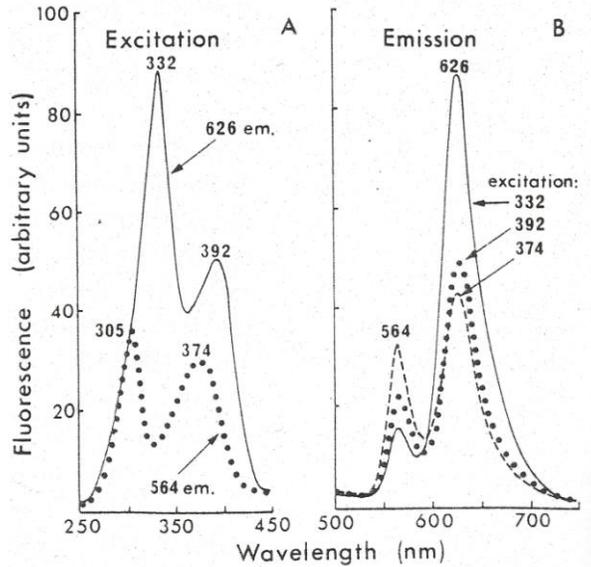

Fig. 8 The fluorescentce spectrum of protein from Malacosteus, compared with spectum of exciting radiation (From A.K. Campbell and P.J. Herring, 1987).

## 6. ASTRONOMICAL RELEVANCE

While the origin of the red rain cells remains uncertain, the possibility of their astronomical relevance has been suggested in several papers (Louis and Kumar, 2003, 2006). In this connection, the hyperthermophile properties discussed in the present paper and the unusual fluorescence behaviour are worthy of note.

We conclude this section by comparing spectra in Fig 7 with astronomical spectra of a fluorescnence phenomenon (ERE emission) for which no convincing abiotic model is still available, Fig 9 shows normalised ERE emission in several astronomical objects and Fig 10 shows the same emission in the famous Red Rectangle, a nebulosity associated with a planetary nebula (Witt and Boronson, 1990; Furton and Witt, 1992, Perrin et al, 1995, Hoyle and Wickramsinghe, 1996). Although non-biological PAH explanations are still being attempted their success has so far been minimal.

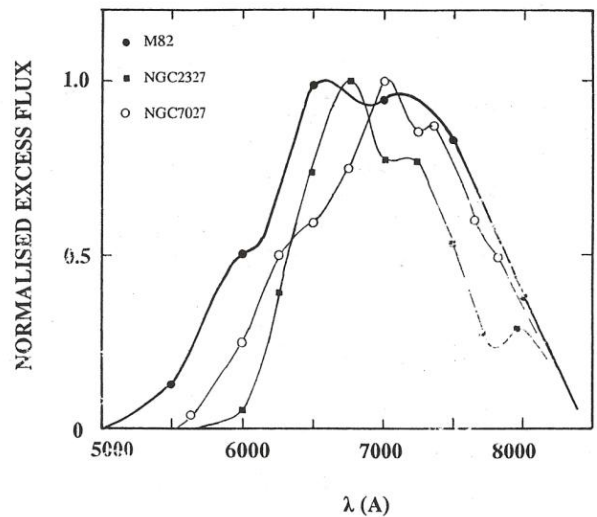

Fig. 9 Normalised excess flux over scattering continuum from data of Furton and Witt (1992) and Perrin et al (1995)

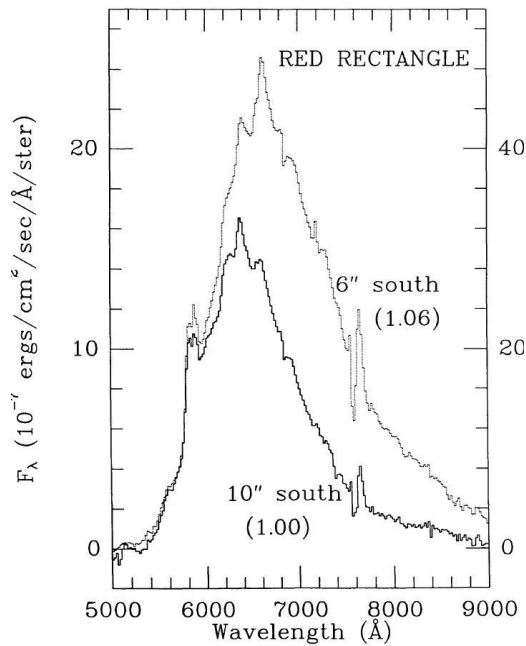

Fig. 10 Normalised excess flux over scattering continuum for the Red Recangle

A spectrum of starlight from a blue star could provide the range of excitaton wavelengths that corresponds to those involved in Fig. 7. The correspondence of profile and peak fluorescence wavelength between the red rain spectra and the ERE spectrum of the red rectangle is impressive. We conclude this paper with a recollection of an earlier comment published by Hoyle and Wickramasinghe:

*"Once again the Universe gives the appearance of being biologically constructed, and on this occasion on a truly vast scale. Once again those who consider such thoughts to be too outlandish to be taken seriously will continue to do so. While we ourselves shall continue to take the view that those who believe they can match the complexities of the Universe by simple experiments in their laboratories will continue to be disappointed."*

## ACKNOWLEDGMENTS

We are grateful to A Hann, M. Thomas and S.Hogg for technical help

## REFERENCES

Baross. A. J and Deming .W. J (1983). Growth of 'Black Smoker' bacteria at termperatures of at least $250^oC$, *Nature*, 303.

Campbell, A. K. and Herring, P. J., (1987) A novel red fluorescent protein from the deep sea luminous fish Malacosteus Niger, *Comp. Biochem. Physiol*, 86B(2), 411-417

Hill.S.C. et al, Fluorescence of Bacteria, Pollen and Naturally occuring Airborne particles: Excitation/Emisson Spectra, Army Research Laboratory, ARL-TR-4722, Feb 2009.

Hoyle, F. and Wickramasinghe, N.C. (1996) Biofluorescence and the Extended Red Emission in astrophysical sources, *Astrophysics and Space Science*, 235, 343-347

Kashefi, K. and Lovely,D.R. (2003). Extending the upper temperature limit for life, *Science*, 301.

Hynum,H.H.,Zeikus,J.S.,Longin,R.,Millet,J. And Ryter,A.(1983). Ultrastructure and extreme heat resistance of spores from thermophilic *Clostridium* species. *Journal of Bacteriology*,15, 1332-1337.

Kentner .D and Sourjik. V, Use of Fluorescence Microscopy to Study Intracellular Signaling in Bacteria, Annual Review of Microbiology, 64, 2010.

Louis, G. and Kumar, A. S. (2006), The red rain phenomenon of Kerala and its possible Extraterretrial Origin, *Astrophysics and Space Science*, 302:175-187.

Louis, G. and Kumar, A. S. (2003) New biology of red rain extremophiles prove cometary panspermia, http://arxiv.org astrophysics e-print archive, arXiv:astro-ph/0312639v1.

Makuch D.S and Irwin. L.N, (2006) The prospect of alien life in exotic forms on other worlds, Naturewissenschaften, 93: 155-172

Perrin, J.M., Darbon, S. and Sivan, J. –P., (1995) CNRS Preprint No 91

Stetter, K.O.(2006). Hypothermophiles in the history of life. *Philosophical Transactions of the Royal Society London, B, Biological Scienes*,361,1837-1843.

Vessoni Penna, T. C., Macheshvili, I.A. and Aquarone,E (1996) *Bacillus stereothermophilus* spores on strips. *Applied Environmental Biotechnology* 40,340-346.

Wainwright, M. (2003). A microbiologist looks at panspermia. *Astrophysics and Space Science,*285,563-570.

Witt, A. N. and Boroson, T. A., (1990) Spectroscopy of extended red emission in reflection nebulae, *Astrophysical Journal*., 355, 182-189